\documentclass[preprint, superscriptaddress,preprintnumbers,amsmath,amssymb,pra]{revtex4}
\usepackage{graphicx}

\begin{document}

\thispagestyle{empty}

\title{Universal experimental test for the role of free charge
carriers in thermal Casimir effect within a micrometer separation
range}

\author{
G.~Bimonte}
\affiliation{Dipartimento di Fisica E. Pancini, Universit\`{a} di Napoli
Federico II, Complesso Universitario MSA, Via Cintia, I-80126 Napoli, Italy}
\affiliation{INFN Sezione di Napoli, I-80126, Napoli, Italy}

\author{
G.~L.~Klimchitskaya}
\affiliation{Central Astronomical Observatory at Pulkovo of the
Russian Academy of Sciences, Saint Petersburg,
196140, Russia}
\affiliation{Institute of Physics, Nanotechnology and
Telecommunications, Peter the Great Saint Petersburg
Polytechnic University, Saint Petersburg, 195251, Russia}

\author{
V.~M.~Mostepanenko}
\affiliation{Central Astronomical Observatory at Pulkovo of the
Russian Academy of Sciences, Saint Petersburg,
196140, Russia}
\affiliation{Institute of Physics, Nanotechnology and
Telecommunications, Peter the Great Saint Petersburg
Polytechnic University, Saint Petersburg, 195251, Russia}
\affiliation{Kazan Federal University, Kazan, 420008, Russia}

\begin{abstract}
We propose a universal experiment to measure the differential Casimir
force between a Au-coated sphere and two halves of a structured plate
covered with a P-doped Si overlayer. The concentration of free charge
carriers in the overlayer is chosen slightly below the critical one, for which
the phase transition from dielectric to metal occurs. One half of the
structured plate is insulating, while its second half is made of gold.
For the former we consider two  different structures, one consisting of
bulk  high-resistivity Si and the other of a  layer
of SiO${}_2$ followed by bulk high-resistivity Si.
The differential Casimir force is computed within the Lifshitz theory
using four approaches that have been proposed in the literature to account for
the role of free charge
carriers in metallic and dielectric materials interacting with quantum
fluctuations. According to these approaches, Au at low frequencies is
described by either the Drude or the plasma model, whereas the free charge
carriers in dielectric materials at room temperature are either taken
into account or disregarded. It is shown that the values of differential
Casimir forces, computed in the micrometer separation range using these
four approaches, are widely distinct from each other and can be easily
discriminated experimentally. It is shown  that for all approaches the thermal
component of the differential Casimir force is sufficiently large for direct
observation. The possible errors and
uncertainties in the proposed experiment are estimated and its importance
for the theory of quantum fluctuations is discussed.
\end{abstract}

\maketitle

\section{Introduction}

During the last two decades the Casimir force \cite{1} acting between
two closely spaced uncharged material surfaces attracted much experimental
and theoretical attention due to the diverse roles it plays in both
fundamental and applied physics (see the monograph \cite{2} and
the
reviews \cite{3,4,5}). The Casimir effect is an entirely quantum and
(at separations between surfaces exceeding two or three nanometers)
relativistic phenomenon. In the nonrelativistic region the Casimir force
is commonly known as van~der~Waals force \cite{6}.
The Casimir effect is caused
by the zero-point and thermal fluctuations of the electromagnetic field.
Although Casimir \cite{1} calculated the force acting between two ideal metal
planes, Lifshitz \cite{7} developed a general theory of van\ der\ Waals
and Casimir forces between two parallel material plates described by
the respective
frequency-dependent dielectric permittivities. In recent years this
theory has been generalized to the case of arbitrarily shaped interacting
bodies \cite{5}.

In 2000 it has been shown that the Lifshitz theory leads to widely different
predictions for the thermal Casimir force between metallic test bodies
depending on whether the
low-frequency dielectric response of metals is described by either the
lossy Drude \cite{8} or lossless plasma \cite{9} model. Later it was
demonstrated that if the Drude model is used the Casimir entropy calculated
within the Lifshitz theory does not satisfy the third law of thermodynamics
(the Nernst heat theorem) for both nonmagnetic \cite{10,11,12,13} and
magnetic \cite{14} metals with perfect crystal lattices.
If instead the plasma model is used, the Lifshitz theory was found to be in perfect
agreement with the Nernst theorem \cite{10,11,12,13,14}.
By contrast, in the limiting case of large separations between metallic
plates, for which classical statistical physics should be applicable,
a first principle computation based on a microscopic model  was shown in Ref.~\cite{15}
to reproduce the Casimir force   predicted by the Lifshitz theory combined with the Drude model.
The Drude model has been later
 shown to satisfy the
Bohr-van~Leeuwen theorem of classical statistical physics,
which is, however, violated if the plasma model is used
to calculate the Casimir force at large separations \cite{16}.
It was shown also that the difference in theoretical predictions
of the Drude and plasma models could be attributed to the
magnetic interaction among fluctuating Foucault currents
\cite{16a,16b,16c}.

In two series of precise experiments on measuring the Casimir interaction
between metals by means of a micromechanical oscillator \cite{17,18,19,20}
and an atomic force microscope \cite{21,22,23}, theoretical predictions
of the Lifshitz theory using the Drude model dielectric response at low
frequencies have been excluded by the measurement data at up to 99\%
confidence level. The theoretical predictions obtained using the plasma
model turned out to be in agreement with the data at a 90\% confidence
level \cite{24}. All these experiments were performed at separations
below one micrometer, where the difference between the theoretical predictions
using the Drude and plasma models does not exceed a few percent.
The only experiment performed at separations up to $7.3\,\mu$m was
interpreted to be in agreement with theoretical predictions
of the Lifshitz theory combined with the Drude model \cite{25}.
It was noted, however, that in Ref.~\cite{25} the Casimir force was
extracted by means of a fitting procedure from up to an order of
magnitude larger measured force, supposedly originating from electrostatic
patches and  influenced by imperfections, which are unavoidably present
on the surfaces of lenses with centimeter-size radii \cite{26,27}.
Thus, the majority of experiments on measuring the Casimir force
do confirm the lossless
plasma model. Since the relaxation
of free electrons in metals at low frequencies is a much studied fact,
this situation is regarded in the literature as the Casimir puzzle.

The Casimir force acting on dielectric bodies also presents a puzzle.
The problem is that the measurement data of all precise experiments
with dielectric surfaces
are in agreement with theoretical predictions of the Lifshitz theory
only provided that the contribution of free charge carriers
 in the dielectric response (dc conductivity)
 is omitted in computations \cite{28,29,30,31}.
If the dc conductivity of dielectric bodies is taken into account, the
theoretical predictions of the Lifshitz theory are excluded by the
measurement data \cite{28,29,30,31,32}. Similar to the role of
relaxation of free charge carriers in metallic bodies, the influence of
the dc conductivity in the Casimir force on dielectric bodies
is limited to just a few
percent of the measured signal. Theoretically it was proven that
the Lifshitz theory with included dc conductivity of dielectric
materials violates the Nernst heat theorem, while an agreement with
this theorem is restored if the dc conductivity is omitted
\cite{33,34,35,36}. This makes us again to disregard a real physical
phenomenon (here, the conductivity of a dielectric material at nonzero
temperature) to achieve agreement with both  experimental data
and the third law of thermodynamics.

It is worth noting that the impact of both relaxation phenomena in metals and
of the dc
conductivity in dielectrics on the Casimir force at separations below
$1\,\mu$m is always relatively small. Then one may hope that by a proper account
of some background effects due to, e.g., electrostatic patches \cite{37}
or surface roughness \cite{3,37aa} one could bring predictions of the literally
understood Lifshitz theory in agreement with the data.
To invstigate this possibility, Ref.~\cite{37a} proposed to measure the
Casimir force between two aligned sinusoidally corrugated Ni surfaces,
one of which is coated with a thin opaque Au layer having a flat surface.
According to the results obtained, the phase-dependent modulation of the
Casimir force for submicron separations, predicted by the Drude model,
is several orders of magnitude larger than that predicted by the plasma
model. An experiment based on this principle \cite{38} has already measured the
differential force between Au- or Ni-coated spheres and Au and Ni
sectors of a structured disc covered with an Au overlayer at
separations of a few hundreds nanometers. The measurement data
unequivocally ruled out the Drude model and were found to be in good agreement
with the plasma model \cite{38}. This experiment is immune to
electrostatic forces caused by  patch potentials similarly to isoelectronic
differential force measurements searching for Yukawa-type corrections
to Newton's gravitational law \cite{39,40}. It is interesting to point out
 that the Casimir free
energies and pressures of thin metallic films computed using the Drude
and plasma models have also been found to differ by up to a factor of several thousands
\cite{41,42,43}. It is found again that the Nernst heat theorem is violated if the
Drude model is used in computations \cite{44}.

In view of the above considerations it would be clearly
desirable to perform an experimental test for the role of free
charge carriers in the Casimir force in the micrometer separation
range where  thermal effects are most pronounced. Differential
force measurements of this sort have been proposed in Refs.~\cite{44a,44b} for both
nonmagnetic and magnetic metallic test bodies.
The setup considered in Refs.~\cite{44a,44b}  involved a structured plate,
one half of which was made of a metal (Au or Ni) and the other half of  high-resistivity Si.
The  structured plate was covered with a plane-parallel overlayer made of B-doped Si plate in the
metallic state with
 a thickness  of 100 nm.
 This  setup allows for a measurable differential
force between a metal-coated sphere and the two halves of a sample, which is
of a quite different magnitude depending on whether the
metal is described by either the Drude
or the plasma model. The test of Refs.~\cite{44a,44b} is not sensitive,
however, to the dc conductivity of high-resistivity Si.

In this paper, we propose a universal test for the role of free charge
carriers in the Casimir force in the micrometer range of separations.
It follows the main idea of Refs.~\cite{44a,44b}, i.e., it suggests to measure
the differential Casimir force between a Au-coated sphere and the high-resistivity
Si and Au halves of a structured plate. The main difference is, however,
that we now consider a semitransparent overlayer made of P-doped Si with
a concentration of free charge carriers which is only slightly
smaller than the critical concentration for which
the dielectric-to-metal phase transition occurs. This means that the
used overlayer is in a dielectric state although it preserves all experimental
advantages of having rather high electronic conductivity.

We calculate the differential Casimir force in the configuration
including both metallic and dielectric materials
using the following four theoretical
approaches:  in the first one
Au at low frequencies is described by the
plasma model and the conductivity of Si is disregarded;
in the second approach
Au at low frequencies is described by the
plasma model but the conductivity of Si is taken into account;
in the third one
 Au at low frequencies is described by the
Drude model but the conductivity of Si is disregarded;
finally, in the fourth approach
the low-frequency dielectric response of Au is described by the
Drude model and the conductivity of Si is taken into account.
We compute the differential Casimir force using these four approaches and
show that in all of them the obtained results are widely different in the
micrometer separation range and can be easily discriminated using the already
available experimental setup. It is shown that the proposed experiment
allows for a precise measurement of the thermal effect in the Casimir force.
We also propose a modified structure of the plate by adding a layer of
SiO${}_2$, which offers some advantages for the process of plate
preparation and further increases the relative differences between theoretical
predictions made by the four approaches.
Computations of the differential Casimir forces are performed using the
tabulated optical data of all involved materials over the frequency ranges
for which they are available. The estimation of both theoretical and
experimental errors and uncertainties in the proposed experiment
demonstrates its feasibility.

The structure of the paper is as follows. In Sec.~II the principle scheme
of the proposed experiment is outlined and the general formalism is presented.
Section~III contains the results of numerical computations of the differential
Casimir force within the four theoretical approaches. Section~IV presents the
modified experimental scheme and more detailed computational results for
the differential force. In Sec.~V all errors and uncertainties are estimated.
In Sec.~VI the reader will find our conclusions and discussion.

\section{Principal experimental scheme and general formalism}

We consider the configuration of a Au-coated sapphire sphere with
radius $R=150\,\mu$m
 in vacuum, at a separation $a$ from a structured plate at room
temperature $T=300\,$K. The thickness of the Au coating
is assumed to exceed 100\,nm, in such a way that it
is legitimate to consider the sphere as
if it were made entirely of
gold, for the sake of computing the Casimir
force. The structured plate consists of an overlayer made of P-doped Si of
thickness $d=100\,$nm covering two sections one of which is made of
high-resistivity Si, while the other is made of Au (see Fig.~1).
The thickness of both
sections is large enough to consider them as two semispaces.
We consider an
overlayer with rather high electronic conductivity corresponding to the
density of free charge carriers $n=3.5\times 10^{18}\,\mbox{cm}^{-3}$.
This density is, however, slightly smaller than the critical density
$n_{\rm cr}=3.84\times 10^{18}\,\mbox{cm}^{-3}$ at which  the dielectric-to-metal
phase transition occurs \cite{45}. Thus, both the overlayer and the underlying
left section of the plate are made of dielectric materials although with
quite different densities of free charge carriers.

We denote the materials of the sphere, of the overlayer, and of the left
section of the plate as 1, 2, and 3, respectively. Then the material of
the right section of the plate is also denoted as 1.
We denote the vacuum gap
as  material 0. The respective dielectric permittivities at the pure
imaginary Matsubara frequencies are
\begin{equation}
\varepsilon_k(i\xi_l)=\varepsilon_k\left(i\frac{c\zeta_l}{2a}\right)
\equiv\varepsilon_{k,l}.
\label{eq1}
\end{equation}
\noindent
Here, $k=0,\,1,\,2,\,3$, $\xi_l=2\pi k_BTl/\hbar$ with $l=0,\,1,\,2,\ldots$
and $k_B$ being the Boltzmann constant are the Matsubara frequencies, and $\zeta_l$
are the dimensionless Matsubara frequencies.
For a vacuum gap we have $\varepsilon_{0,l}=1$.

In the proposed experiment the sphere moves back and forth with
sufficiently high frequency at some fixed
separation $a$ from the plate. In this case not
 the Casimir forces $F_{\rm Si}(a,T)$
and $F_{\rm Au}(a,T)$
among the Au sphere and the left and right sections of the plate
but only the difference $F_{\rm diff}(a,T)$ between them
is measured \cite{38,39,40}:
\begin{equation}
F_{\rm diff}(a,T)=F_{\rm Si}(a,T)-F_{\rm Au}(a,T)
\label{eq2}
\end{equation}
\noindent
This measurement should be repeated at different separations.
Note that $F_{\rm Si}(a,T)$ and $F_{\rm Au}(a,T)$ are the forces acting
when the sphere bottom is above some points deep in the left and right
halves of the plate, respectively. Because of this, one can neglect the
effect of sharp boundary between the left and right
halves of the plate and consider each of them as infinitely
large \cite{44b}.

Using the proximity force approximation \cite{2}, which was recently shown
to be sufficiently exact under the condition $a\ll R$ \cite{46,47,48}, and
the Lifshitz formula \cite{2,7}, the differential force (\ref{eq2}) can be
calculated by the following equation:
\begin{eqnarray}
&&
F_{\rm diff}(a,T)=\frac{k_BTR}{4a^2}
\sum_{l=0}^{\infty}{\vphantom{\sum}}^{\prime}
\int_{\zeta_l}^{\infty}\!\!ydy
\label{eq3} \\
&&~~~\times
\sum_{\alpha}
\ln\frac{1-r_{\alpha}^{(0,1)}(i\zeta_l,y)R_{\alpha}^{(0,2,3)}(i\zeta_l,y)
e^{-y}}{1-r_{\alpha}^{(0,1)}(i\zeta_l,y)R_{\alpha}^{(0,2,1)}(i\zeta_l,y)
e^{-y}},
\nonumber
\end{eqnarray}
\noindent
where the prime on the summation sign denotes that the $l=0$
term is taken with weight 1/2.
Here, $y$ is the dimensionless variable connected with
the projection of the wave vector
onto the plane of the plate, $\mbox{\boldmath{$k$}}_\bot$, by
\begin{equation}
y=2a\sqrt{\mbox{\boldmath{$k$}}_{\bot}^2+\frac{\xi_l^2}{c^2}}.
\label{eq4}
\end{equation}
\noindent
The summation in $\alpha$ is made over the two independent polarizations
of the electromagnetic field, transverse magnetic ($\alpha=\mbox{TM}$) and
transverse electric ($\alpha=\mbox{TE}$). The reflection coefficients on the
structured plate covered with the overlayer are expressed as
\begin{equation}
R_{\alpha}^{(0,2,j)}(i\zeta_l,y)=
\frac{r_{\alpha}^{(0,2)}(i\zeta_l,y)+r_{\alpha}^{(2,j)}(i\zeta_l,y)
e^{-d\sqrt{y^2+(\varepsilon_{2,l}-1)\zeta_l^2}/a}}{1+
r_{\alpha}^{(0,2)}(i\zeta_l,y)r_{\alpha}^{(2,j)}(i\zeta_l,y)
e^{-d\sqrt{y^2+(\varepsilon_{2,l}-1)\zeta_l^2}/a}},
\label{eq5}
\end{equation}
\noindent
where $j=1,\,3$. Finally, the reflection coefficients on the boundary
surfaces between two different materials are given by
the familiar Fresnel formulas taken at the imaginary Matsubara
frequencies:
\begin{eqnarray}
&&
r_{\rm TM}^{(k,j)}(i\zeta_l,y)=\frac{\varepsilon_{j,l}
\sqrt{y^2+(\varepsilon_{k,l}-1)\zeta_l^2}-\varepsilon_{k,l}
\sqrt{y^2+(\varepsilon_{j,l}-1)\zeta_l^2}}{\varepsilon_{j,l}
\sqrt{y^2+(\varepsilon_{k,l}-1)\zeta_l^2}+\varepsilon_{k,l}
\sqrt{y^2+(\varepsilon_{j,l}-1)\zeta_l^2}},
\nonumber \\
&&
r_{\rm TE}^{(k,j)}(i\zeta_l,y)=\frac{\sqrt{y^2+(\varepsilon_{k,l}-1)\zeta_l^2}-
\sqrt{y^2+(\varepsilon_{j,l}-1)\zeta_l^2}}{\sqrt{y^2+(\varepsilon_{k,l}-1)\zeta_l^2}+
\sqrt{y^2+(\varepsilon_{j,l}-1)\zeta_l^2}},
\label{eq6}
\end{eqnarray}
where $(k,j)=(0,1),\,(0,2),\,(2,1)$, and (2,3).

The contribution to Eq.~(\ref{eq3}) at zero Matsubara frequency requires special
attention because the problems discussed in Sec.~I are connected with different
treatments of this contribution. The point is that the optical data for the
complex index of refraction of all materials
are available only for sufficiently large frequencies.
Therefore, at low frequencies the available data should be supplemented with
some theoretical model. Taking into account the relaxation properties of electrons
in metals, the dielectric permittivities of our material 1 (Au) at the Matsubara
frequencies can be represented in the form
\begin{equation}
\varepsilon_{1,l}^{D}=
\frac{\tilde{\omega}_{p,1}^2}{\zeta_l(\zeta_l+\tilde{\gamma_1})}
+\varepsilon_{1,l}^{\rm cor}.
\label{eq7}
\end{equation}
\noindent
Here, $\tilde{\omega}_{p,1}$ and $\tilde{\gamma}_1$ are,
respectively, the dimensionless plasma
frequency and relaxation parameter of Au connected with the
dimensional ones by
\begin{equation}
\tilde{\omega}_{p,1}=\frac{2a\omega_{p,1}}{c},\quad
\tilde{\gamma}_{1}=\frac{2a\gamma_{1}}{c}
\label{eq8}
\end{equation}
\noindent
and $\varepsilon_{1,l}^{\rm cor}$ is a contribution of the core (bound)
electrons to the dielectric permittivity determined by the optical data.
The upper index $D$ is used to underline that the permittivity (\ref{eq7})
has the Drude form. Note that the relaxation parameter $\tilde{\gamma}_1$
depends on temperature and goes to zero with vanishing $T$ by a power law.
For metals with disregarded relaxation properties of free electrons the
dielectric permittivity takes the plasma form
\begin{equation}
\varepsilon_{1,l}^{p}=
\frac{\tilde{\omega}_{p,1}^2}{\zeta_l^2}
+\varepsilon_{1,l}^{\rm cor}.
\label{eq9}
\end{equation}
\noindent
The plasma model is usually used in the region of infrared optics where
$\gamma_1\ll\xi_l$.

For dielectric materials ($k=2,\,3$) with free charge
carriers taken into account
the dielectric permittivity at the Matsubara frequencies takes the form
\begin{equation}
\varepsilon_{k,l}=
\frac{4\pi\tilde{\sigma}_{k,l}}{\zeta_l}
+\varepsilon_{k,l}^{\rm cor},
\label{eq10}
\end{equation}
\noindent
where the dimensionless conductivity $\tilde{\sigma}_{k,l}$ is connected with
the dimensional one by
\begin{equation}
\tilde{\sigma}_{k,l}=\frac{2a\sigma_{k,l}}{c}.
\label{eq11}
\end{equation}
\noindent
The conductivity of dielectrics is temperature-dependent. It vanishes
exponentially fast when $T$
goes to zero. At room temperature it is customary to represent the frequency
dependence of $\tilde{\sigma}_{k,l}$ in terms of conventional Drude
parameters, i.e., as
\begin{equation}
\tilde{\sigma}_{k,l}=
\frac{\tilde{\omega}_{p,k}^2}{4\pi(\zeta_l+\tilde{\gamma_k})}.
\label{eq11a}
\end{equation}
\noindent
If the free charge carriers in dielectric materials are disregarded, it holds
\begin{equation}
\varepsilon_{k,l}=
\varepsilon_{k,l}^{\rm cor},
\quad k=2,\,3.
\label{eq12}
\end{equation}
\noindent
For our dielectric materials with $k=2$ (P-doped Si) and $k=3$
(high-resistivity Si)
\begin{equation}
\varepsilon_{2,l}^{\rm cor}=
\varepsilon_{3,l}^{\rm cor},
\label{eq13}
\end{equation}
\noindent
although $\tilde{\sigma}_{2,l}$ and $\tilde{\sigma}_{3,l}$  are significantly
different. Note also that for both metals and dielectrics
$\varepsilon_{k,l}^{\rm cor}\to 1$ when $\zeta_l\to\infty$.

At first, we assume that the free charge carriers in both dielectric materials
(i.e., in
the overlayer made of doped Si and in the plate section of high-resistivity Si) are
disregarded. Then, at $l=0$ Eqs.~(\ref{eq12}) and (\ref{eq13}) hold.
In this case, from the first line of Eq.~(\ref{eq6}) one obtains
$r_{\rm TM}^{(2,3)}(0,y)=0$ and from  Eq.~(\ref{eq5}) we have
\begin{equation}
R_{\rm TM}^{(0,2,3)}(0,y)=r_{\rm TM,0}^{(0,2)}=
\frac{\varepsilon_{2,0}^{\rm cor}-1}{\varepsilon_{2,0}^{\rm cor}+1}.
\label{eq14}
\end{equation}
\noindent
Now, from the first line of Eq.~(\ref{eq6}), one can see that
$r_{\rm TM}^{(2,1)}(0,y)=1$ independently of whether Au is described by the
Drude model of Eq.~(\ref{eq7}) or by the plasma model of Eq.~(\ref{eq9}).
Then, Eq.~(\ref{eq5}) leads to
\begin{equation}
R_{\rm TM}^{(0,2,1)}(0,y)=
\frac{r_{\rm TM,0}^{(0,2)}+e^{-dy/a}}{1+r_{\rm TM,0}^{(0,2)}e^{-dy/a}},
\label{eq15}
\end{equation}
\noindent
where $r_{\rm TM,0}^{(0,2)}$ is given by Eq.~(\ref{eq14}).

{}From the second line of Eq.~(\ref{eq6}) we have
\begin{equation}
r_{\rm TE}^{(0,2)}(0,y)=r_{\rm TE}^{(2,3)}(0,y)=0.
\label{eq16}
\end{equation}
\noindent
Then, from Eq.~(\ref{eq5}) one obtains
\begin{equation}
R_{\rm TE}^{(0,2,3)}(0,y)=0.
\label{eq17}
\end{equation}
\noindent
As to the reflection coefficient $R_{\rm TE}^{(0,2,1)}(0,y)$, its value
depends on whether Au is described by the Drude or the plasma model.
If the Drude model (\ref{eq7}) is used, the second line of Eq.~(\ref{eq6})
leads to $r_{{\rm TE},D}^{(2,1)}(0,y)=0$ and Eq.~(\ref{eq5}) results in
\begin{equation}
R_{{\rm TE},D}^{(0,2,1)}(0,y)=0,
\label{eq18}
\end{equation}
\noindent
similar to Eq.~(\ref{eq17}). If, however, the plasma model (\ref{eq9}) is
used for Au, the second line of Eq.~(\ref{eq6})
leads to
\begin{equation}
r_{{\rm TE},p}^{(2,1)}(0,y)=
\frac{y-\sqrt{y^2+\tilde{\omega}_{p,1}^2}}{y+
\sqrt{y^2+\tilde{\omega}_{p,1}^2}}
\label{eq19}
\end{equation}
\noindent
and from Eq.~(\ref{eq5}) one obtains
\begin{equation}
R_{{\rm TE},p}^{(0,2,1)}(0,y)=
\frac{y-\sqrt{y^2+\tilde{\omega}_{p,1}^2}}{y+
\sqrt{y^2+\tilde{\omega}_{p,1}^2}}\,e^{-dy/a}.
\label{eq20}
\end{equation}

Next, we assume an alternative assumption that the free charge carriers
in dielectric materials are taken into account by Eq.~(\ref{eq10}).
In this case $r_{\rm TM}^{(0.2)}(0,y)=1$ and from Eq.~(\ref{eq5}) we have
\begin{equation}
R_{\rm TM}^{(0,2,3)}(0,y)=R_{\rm TM}^{(0,2,1)}(0,y)=
1.
\label{eq21}
\end{equation}
\noindent
Notice that the latter equation is valid irrespective of whether
Au is described by the Drude or
the plasma models. For the TE mode Eq.~(\ref{eq16}) is still valid leading to
\begin{equation}
R_{\rm TE}^{(0,2,3)}(0,y)=0.
\label{eq22}
\end{equation}
\noindent
The value of the coefficient $R_{\rm TE}^{(0,2,1)}(0,y)$ depends on the model
used for a description of Au. If the Drude model is used,
\begin{equation}
r_{{\rm TE},D}^{(2,1)}(0,y)=R_{{\rm TE},D}^{(0,2,1)}(0,y)=0.
\label{eq23}
\end{equation}
\noindent
If, however, Au is described by the plasma model, one returns back to
Eqs.~(\ref{eq19}) and (\ref{eq20}), which preserve their validity when the
free charge carriers in dielectric materials are taken into account.

{}From the above one can see that at zero Matsubara frequency the coinciding
reflection coefficients $R_{\rm TE}^{(0,2,j)}$ with $j=1,\,3$ are obtained
in the cases when the free charge carriers of the dielectric materials 2 and 3
are either disregarded or taken into account. In each case, however, the
results for $R_{\rm TE}^{(0,2,1)}$ depend on the model used for a description
of Au. Just the opposite situation
 holds for the coefficients $R_{\rm TM}^{(0,2,j)}$ calculated at $\zeta_0=0$.
 Here, the results do not depend on the model of a metal, but are different
 depending on whether or not the free charge carriers of dielectric materials
 are taken into account.

\section{Discrimination between different theoretical approaches}

Now we perform numerical computations of the differential Casimir force (\ref{eq2})
in the configuration of Sec.~II by using Eqs.~(\ref{eq3})--(\ref{eq5})
and the
explicit expressions for the reflection coefficients at zero Matsubara frequency
provided in Eqs.~(\ref{eq14}),  (\ref{eq15}), (\ref{eq17}), (\ref{eq18}),
(\ref{eq20})--(\ref{eq23}) obtained in the framework of different theoretical
approaches. The dielectric permittivities $\varepsilon_{k,l}^{\rm cor}$ of Au
and Si at the Matsubara frequencies were obtained using the tabulated optical data
for the complex index of refraction \cite{49,50} and the Kramers-Kronig
relation following Refs.~\cite{2,3}. For Au the values of the plasma frequency
$\omega_{p,1}\approx 9\,\mbox{eV}=1.37\times 10^{16}\,$rad/s
and the relaxation parameter
$\gamma_{1}\approx 35\,\mbox{meV}=5.3\times 10^{13}\,$rad/s
have been used.

For P-doped Si the chosen concentration of free electrons
($n=3.5\times 10^{18}\,\mbox{cm}^{-3}$) corresponds to the plasma frequency
\begin{equation}\omega_{p,2}=e\sqrt{\frac{4\pi n}{m^{\ast}}}\approx
2.1\times 10^{14}\,\mbox{rad/s},
\label{eq24}
\end{equation}
\noindent
where the effective electron mass is $m^{\ast}=0.26m_e$.
The chosen value of $n$ corresponds also to the resistivity
$\rho_2=(1.4\pm 0.1)\times 10^{-2}\,\Omega\,$cm \cite{51}, i.e.,
$\rho_2\approx 1.55\times 10^{-14}\,$s and conductivity
$\sigma_2\approx 0.64\times 10^{14}\,\mbox{s}^{-1}$.
If one models the conductivity of P-doped Si at fixed $T=300\,$K
by means of the Drude model, this value of conductivity leads to the
following relaxation parameter \cite{52}
\begin{equation}
\gamma_2=\frac{\omega_{p,2}^2}{4\pi\sigma_2}=
\frac{1}{4\pi}\rho_2\omega_{p,2}^2\approx 5.5\times 10^{13}\,
\mbox{rad/s}.
\label{eq25}
\end{equation}

Now we consider the material 3, i.e., high-resistivity Si  whose
conductivity is by about five orders of magnitude lower than that
of our P-doped Si. If the free charge carriers in Si are disregarded,
 materials 2 and 3 become identical and their common dielectric permittivity
is given by $\varepsilon_{2,l}^{\rm cor}$. Note, that for Si
$\varepsilon_{2,0}^{\rm cor}\approx 11.67$.
Inclusion of free charge carriers for  high-resistivity Si affects
the dielectric permittivity only at the zero Matsubara frequency and
results in nonzero dc conductivity. Now we take into account that according
to Eqs.~(\ref{eq17}) and (\ref{eq22}) the reflection coefficient
$R_{\rm TE}^{(0,2,3)}(0,y)$
takes the same value irrespective of whether one includes or
neglects
the contribution to the permittivity of
free charge carriers in Si.
On the contrary the value of the coefficient  $R_{\rm TM}^{(0,2,3)}(0,y)$
does depend on whether one includes or not
the contribution of free charge carriers in the P-doped
Si overlayer [see Eqs.~(\ref{eq14}) and (\ref{eq21})].
However, even in this case it does not depend
on the conductivity properties of high-resistivity Si.

The results of the computations for the differential Casimir force (\ref{eq2}) at
$T=300\,$K are presented in Fig.~\ref{fg2} as functions of separation in the
range from 0.5 to $3.5\,\mu$m and, on an enlarged scale, in an inset in the
range from 3 to $5\,\mu$m. The pair of top lines (solid and dashed) is
computed using the plasma model (\ref{eq9}) for Au. The free charge carriers
in both the P-doped and high-resistivity Si are either disregarded according
to Eq.~(\ref{eq12}) (the solid line) or taken into account according
to Eq.~(\ref{eq10}) (the dashed line). In a similar way, the pair of solid
and dashed bottom lines is
computed using the Drude model (\ref{eq7}) for Au.
Here, the solid line is again computed by disregarding the free charge carriers
in dielectric materials and the dashed line takes these charge carriers into
account.

As it is seen in Fig.~\ref{fg2}, the four theoretical approaches discussed
in Sec.~I lead to widely different predictions for the differential Casimir
force. They can be easily discriminated experimentally keeping in mind that
the sensitivity of difference force measurements of this type is equal to
1\,fN \cite{38} or even a fraction of 1\,fN \cite{40}. Thus, for $a=1\,\mu$m
 use of plasma model for Au results in
$F_{\rm diff}^{p}\approx 138.54\,$fN and
$F_{\rm diff}^{p,dc}\approx 120.66\,$fN
with disregarded and included free charge carriers, respectively, in both
P-doped Si overlayer and high-resistivity Si (the top pair of lines).
If Au is described by the Drude model and the free charge carriers are either
disregarded or included one has
$F_{\rm diff}^{D}\approx 66.37\,$fN or
$F_{\rm diff}^{D,dc}\approx 48.50\,$fN, respectively
(the bottom pair of lines).
It is seen that the four theoretical predictions are approximately 18, 54
and 18\,fN apart, i.e., the force intervals between them far exceed the
experimental sensitivity. Even for $a=2\,\mu$m we have
$F_{\rm diff}^{p}\approx 28.07\,$fN and
$F_{\rm diff}^{p,dc}\approx 23.63\,$fN with disregarded and taken into account
free charge carriers in Si materials, respectively, and
$F_{\rm diff}^{D}\approx 7.71\,$fN and
$F_{\rm diff}^{D,dc}\approx 3.26\,$fN under the same assumptions about the
free charge carriers. Here, the theoretical predictions are approximately 5,
16, and 4\,fN apart, i.e., again can be experimentally discriminated
(see Sec.V for additional information about errors and uncertainties in this
experiment).

In the above computations, the P-doped Si overlayer of $d=100\,$nm thickness
has been used. It is interesting to determine the dependence of $F_{\rm diff}$
on $d$. In Fig.~\ref{fg3} the computational results for the differential
Casimir force at $a=1\,\mu$m, $T=300\,$K are presented as a function of overlayer
thickness using the four theoretical approaches described above (the top pair of the
solid and dashed lines is computed using the plasma model for Au with disregarded
and included free charge carriers in Si, respectively;  the bottom pair of the
solid and dashed lines is computed using the Drude model with either disregarded
or included free charge carriers in Si). As it can be seen in Fig.~\ref{fg3}, the
differential Casimir force decreases monotonously with increasing $d$,
while the discrepancies
 between the theoretical predictions of the different models are almost
 independent on the thickness.
This feature of the force
makes a thicker overlayer  (for instance, $d=200\,$nm, see Sec.~IV)
preferable because in this case the relative error in the determination of $d$
becomes negligibly small and therefore it
does not influence the value of $F_{\rm diff}$.
A more detailed analysis of theoretical errors is contained in Sec.~V.

Now we show that the proposed experiment not only allows
for an easy discrimination between
the four theoretical approaches described above, but it can be used to measure the
thermal effect in $F_{\rm diff}$ as well. First, we illustrate this statement for the
case of Au described by either the Drude or the plasma model with disregarded
free charge carriers in Si. In Fig.~\ref{fg4} the bottom (Drude model) and the
top (plasma model) solid lines are reproduced from Fig.~\ref{fg2} using
a
logarithmic scale along the axis of $F_{\rm diff}$. The middle line in
Fig.~\ref{fg4} shows the computational results for $F_{\rm diff}$ obtained at
$T=0$K. For perfect crystal lattices the relaxation parameter $\gamma(T=0)=0$,
so that the theoretical results obtained using the Drude and plasma models
coincide. For real metals, however, there is some small residual relaxation
$\gamma_{\rm res}(T=0)\neq 0$. We have used
$\gamma_{\rm res,1}=1.2\times 10^{-6}\,$eV for Au and obtained the same middle line
in Fig.~\ref{fg4} as given by the plasma model. Numerically, the computational
results turned out to be very close. For example, at $a=0.5$, 1.0 and $1.5\mu$m
the differential Casimir forces computed at $T=0$ by using the Drude and the plasma
models are equal to 643.11, 643.86\,fN; 124.96, 125.19\,fN; and 43.64, 43.76\,fN, respectively.

As it can be seen in Fig.~\ref{fg4}, at separations $a=0.5$, 1.0, 1.5, 2.0, and
$2.5\mu$m the thermal correction in $F_{\rm diff}$
\begin{equation}
\Delta_{T}F_{\rm diff}(a,T)=F_{\rm diff}(a,T)-F_{\rm diff}(a,0)
\label{eq26a}
\end{equation}
\noindent
computed using the Drude model
(i.e., the difference between the bottom and middle lines) is equal to
--213.28, --58.59, --24.50, --12.34, and --6.5\,fN, respectively.
Note that the quantity $F_{\rm diff}(a,0)$ in Eq.~(\ref{eq26a}) is
computed by the zero-temperature Lifshitz formula \cite{2,7}
where one makes an integration over the continuous frequency
$\zeta$ instead of a summation over the discrete Matsubara
frequencies $\zeta_l$ and uses the zero-temperature values of
all involved dielectric permittivities.
The thermal correction in $F_{\rm diff}$ computed using the plasma model
(i.e., the difference between the top and middle lines) is equal to
18.3, 13.35, 10.11, 7.95, and 6.42\,fN.
In all these cases the magnitudes of the thermal effect are far
larger than the
experimental sensitivity.

In Fig.~\ref{fg4a} we present the computational results for the
thermal correction (\ref{eq26a})  as functions of
separation
at $T=300\,$K using the four theoretical approaches
described above. The top pair of solid and dashed lines is
computed using the plasma model for Au with omitted and
included free charge carriers of dielectrics, respectively.
The bottom pair of solid and dashed lines is also computed
by including and disregarding the free charge carriers in
dielectrics, but describing Au by the Drude model.
The thermal corrections presented by the top and bottom
solid lines (the plasma and Drude models for Au with
omitted charge carriers in dielectrics) have been already
discussed above on the basis of Fig.~\ref{fg4}.
The thermal correction shown by the bottom dashed line
(the Drude model for Au with included free
charge carriers in dielectrics) can be easily observed
over the separation range from 0.5 to $2.5\,\mu$m and
discriminated from the bottom solid line.
As to the thermal correction shown by the top dashed
line (the plasma model for Au with included free
charge carriers in dielectrics), it can be observed
within the separation range from 0.5 to $0.9\,\mu$m.

It is instructive to separate the thermal correction
to the differential force (\ref{eq26a}) into two parts
\begin{equation}
\Delta_{T}F_{\rm diff}(a,T)=\Delta_{T}^{\!(1)}F_{\rm diff}(a,T)
+\Delta_{T}^{\!(2)}F_{\rm diff}(a,T).
\label{eq26b}
\end{equation}
\noindent
Here, we have introduced the notations
\begin{eqnarray}
&&
\Delta_{T}^{\!(1)}F_{\rm diff}(a,T)=\tilde{F}_{\rm diff}(a,T)
-F_{\rm diff}(a,0),
\nonumber \\
&&
\Delta_{T}^{\!(2)}F_{\rm diff}(a,T)=F_{\rm diff}(a,T)
-\tilde{F}_{\rm diff}(a,T),
\label{eq26c}
\end{eqnarray}
\noindent
where $\tilde{F}_{\rm diff}(a,T)$ is calculated by Eq.~(\ref{eq3})
 at $T=300\,$K, but with the zero-temperature values of all
dielectric permittivities. From Eq.~(\ref{eq26c}) it is clear
that $\Delta_{T}^{\!(1)}F_{\rm diff}$ represents
the contribution to the thermal
correction caused by a summation over the discrete Matsubara
frequencies, whereas $\Delta_{T}^{\!(2)}F_{\rm diff}$
originates from the explicit dependence of dielectric
permittivities on the temperature as a parameter.

At the end of this section, we briefly discuss the role
of each  of the two contributions to the thermal correction in the
four theoretical models. If Au is described by the plasma
model and free charge carriers in dielectrics are disregarded,
it holds
\begin{equation}
\Delta_{T}F_{\rm diff}^{p}(a,T)=
\Delta_{T}^{\!(1)}F_{\rm diff}^{p}(a,T)>0
\label{eq26d}
\end{equation}
\noindent
because in this case all dielectric permittivities are
temperature-independent.

If Au is described by the plasma
model and free charge carriers in dielectrics are taken
into account, we have
\begin{eqnarray}
&&
\Delta_{T}^{\!(1)}F_{\rm diff}^{p,dc}(a,T)=
\Delta_{T}{F}_{\rm diff}^{p}(a,T),
\nonumber \\
&&
\Delta_{T}^{\!(2)}F_{\rm diff}^{p,dc}(a,T)<0.
\label{eq26e}
\end{eqnarray}
\noindent
Note that the correction $\Delta_{T}^{\!(2)}F_{\rm diff}^{p,dc}$
is given by the difference between dashed and solid top lines in
Fig.~\ref{fg4a}. For example, at separations of 0.5 and $1\,\mu$m
$\Delta_{T}^{\!(2)}F_{\rm diff}^{p,dc}$ is equal to --64.1 and
--17.1\,fN, respectively.

If Au is desctibed by the Drude
model and the free charge carriers in dielectrics are disregarded,
the dominant contribution to $\Delta_{T}F_{\rm diff}^{D}$ is given by
\begin{equation}
\Delta_{T}^{\!(1)}F_{\rm diff}^{D}(a,T)<0.
\label{eq26f}
\end{equation}
\noindent
The much smaller thermal correction
$\Delta_{T}^{\!(2)}F_{\rm diff}^{D}$ originates from
the difference between the values of the relaxation parameter
of Au at $T=0$ and $300\,$K. As an example,  one has
$\Delta_{T}^{\!(1)}F_{\rm diff}^{D}=-207.0\,$fN,
$\Delta_{T}^{\!(2)}F_{\rm diff}^{D}=-7.0\,$fN at $a=0.5\,\mu$m  and
$\Delta_{T}^{\!(1)}F_{\rm diff}^{D}=-58.0\,$fN,
$\Delta_{T}^{\!(2)}F_{\rm diff}^{D}=-0.8\,$fN at $a=1\,\mu$m.

Finally, if  Au is described by the Drude
model and the free charge carriers in dielectrics are
taken into account, one obtains
\begin{eqnarray}
&&
\Delta_{T}^{\!(1)}F_{\rm diff}^{D,dc}(a,T)=
\Delta_{T}^{\!(1)}{F}_{\rm diff}^{D}(a,T),
\nonumber \\
&&
\Delta_{T}^{\!(2)}F_{\rm diff}^{p,dc}(a,T)<0.
\label{eq26g}
\end{eqnarray}
\noindent
The latter correction contributes significantly to the total
thermal correction. For example, at $a=0.5$ and $1\,\mu$m
$\Delta_{T}^{\!(2)}F_{\rm diff}^{D,dc}$ is equal to
--71.1 and --18.0\,fN.

\section{Experimental test with additional silica layer}

\begin{figure}[b]
\vspace*{-5cm}
\centerline{\hspace*{3.5cm}
\includegraphics{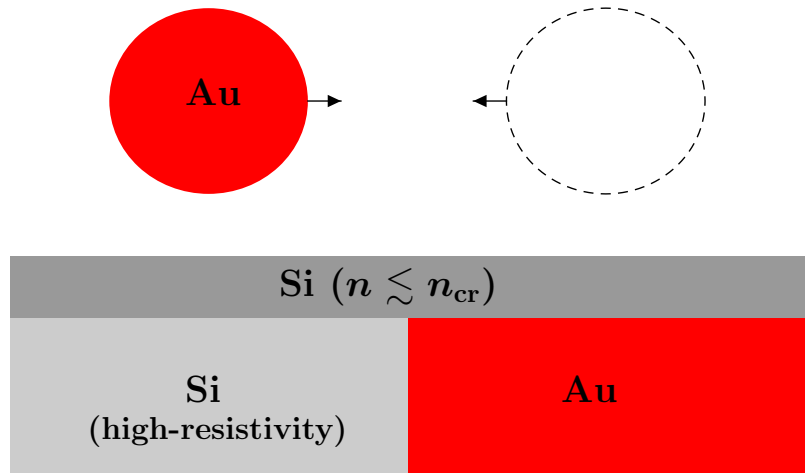}
}
\vspace*{-12cm}
\caption{\label{fg1}
The experimental configuration of a Au sphere moving back and
forth above a structured
plate covered with a P-doped Si overlayer in the dielectric state.
The measured quantity is the differential Casimir force $F_{\rm diff}$
between the Au sphere and the two halves of the plate when the sphere bottom
is far away from their boundaries. The figure displays the two extreme
positions of the sphere during its motion.
The size of the sphere is shown not to scale.
}
\end{figure}

\begin{figure}[b]
\vspace*{-2cm}
\centerline{\hspace*{3.5cm}
\includegraphics{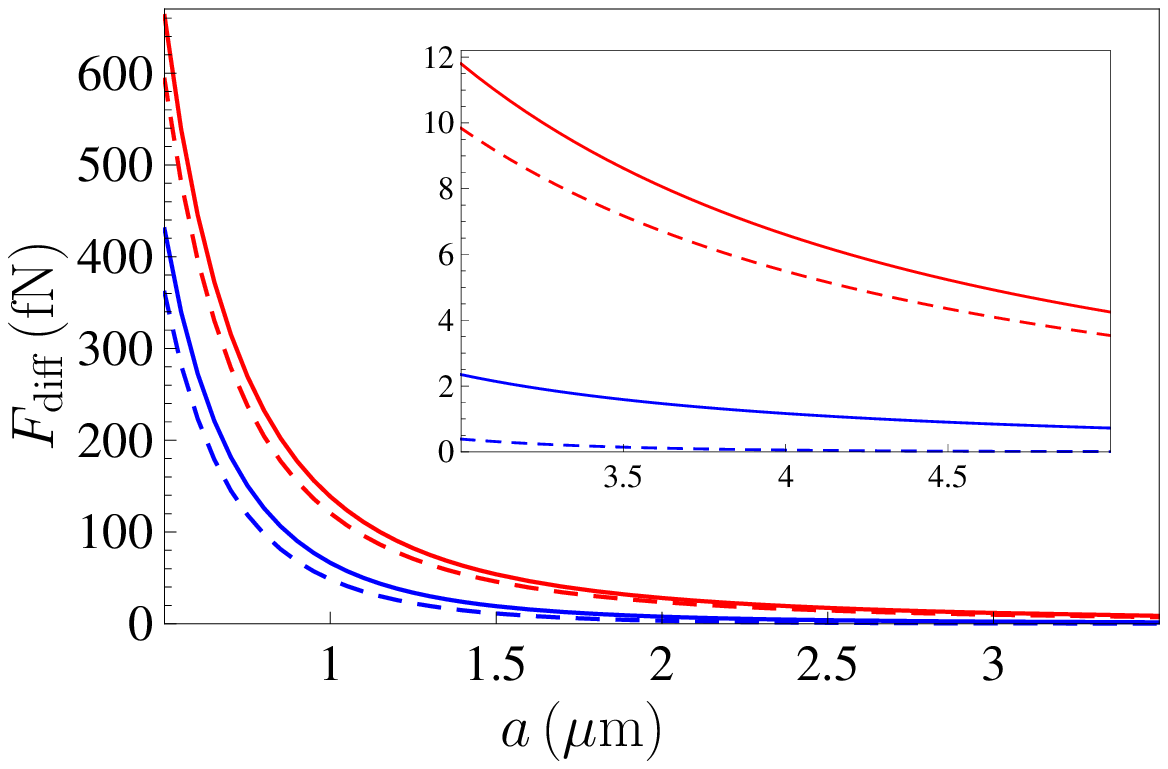}
}
\vspace*{-10cm}
\caption{\label{fg2}
Differential Casimir forces computed at $T=300\,$K
in the configuration of Fig.~\ref{fg1}
using the plasma model for Au with disregarded and taken into account
free charge carriers in dielectric materials (top pair of solid
and dashed lines, respectively) and
using the Drude model for Au with disregarded and included
free charge carriers in dielectrics (bottom pair of solid
and dashed lines, respectively) are shown as the functions of
separation. The region of larger distances is displayed in an inset.
}
\end{figure}

\begin{figure}[b]
\vspace*{-7cm}
\centerline{\hspace*{3.5cm}
\includegraphics{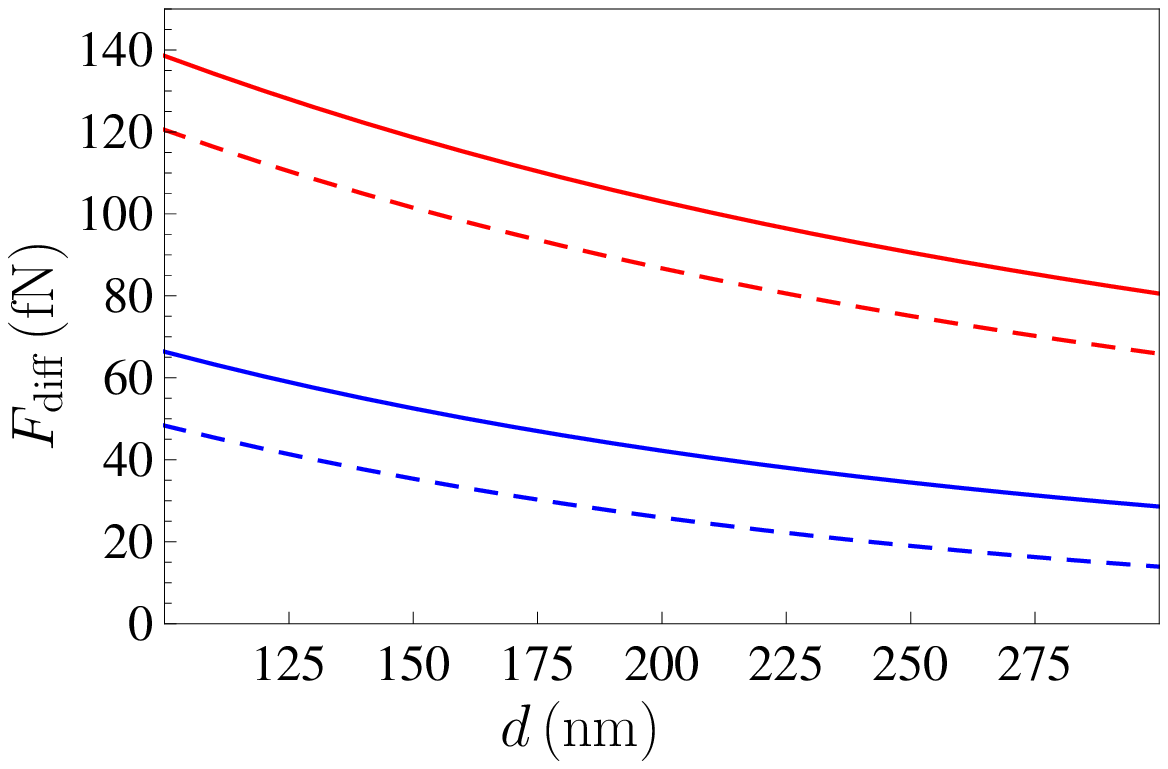}
}
\vspace*{-10cm}
\caption{\label{fg3}
Differential Casimir forces computed at $T=300\,$K
in the configuration of Fig.~\ref{fg1}
using the plasma model for Au with disregarded and taken into account
free charge carriers in dielectric materials (top pair of solid
and dashed lines, respectively) and
using the Drude model for Au with disregarded and included
free charge carriers in dielectrics (bottom pair of solid
and dashed lines, respectively) are shown as the functions of
overlayer thickness at $a=1\,\mu$m.
}
\end{figure}

\begin{figure}[b]
\vspace*{-7cm}
\centerline{\hspace*{3.5cm}
\includegraphics{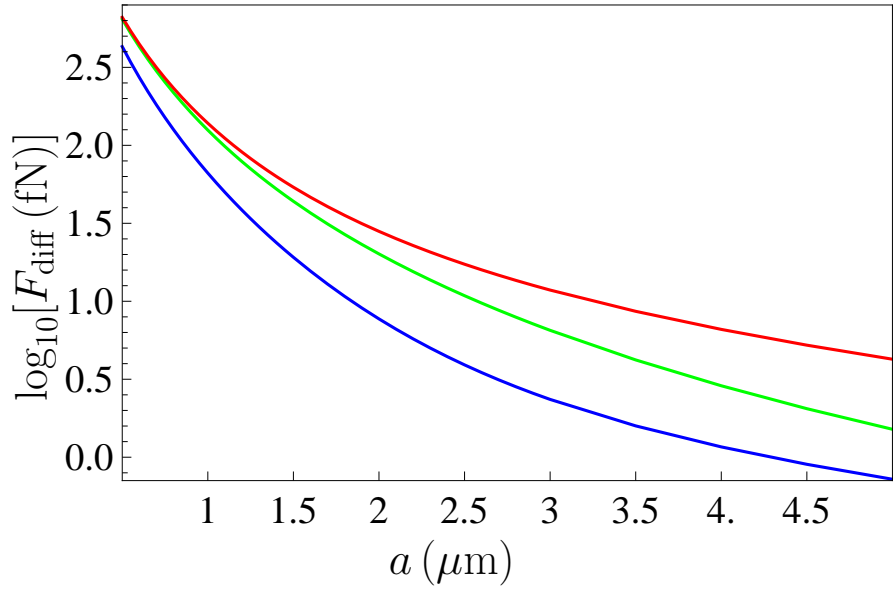}
}
\vspace*{-10cm}
\caption{\label{fg4}
Logarithm of the differential Casimir forces computed  at $T=300\,$K
using the plasma and Drude models for Au with disregarded
free charge carriers in dielectric materials are shown as functions of
separation by the top and bottom lines, respectively.
The medium line shows common computational results for the differential
Casimir force at zero temperature.
}
\end{figure}

\begin{figure}[b]
\vspace*{-7cm}
\centerline{\hspace*{3.5cm}
\includegraphics{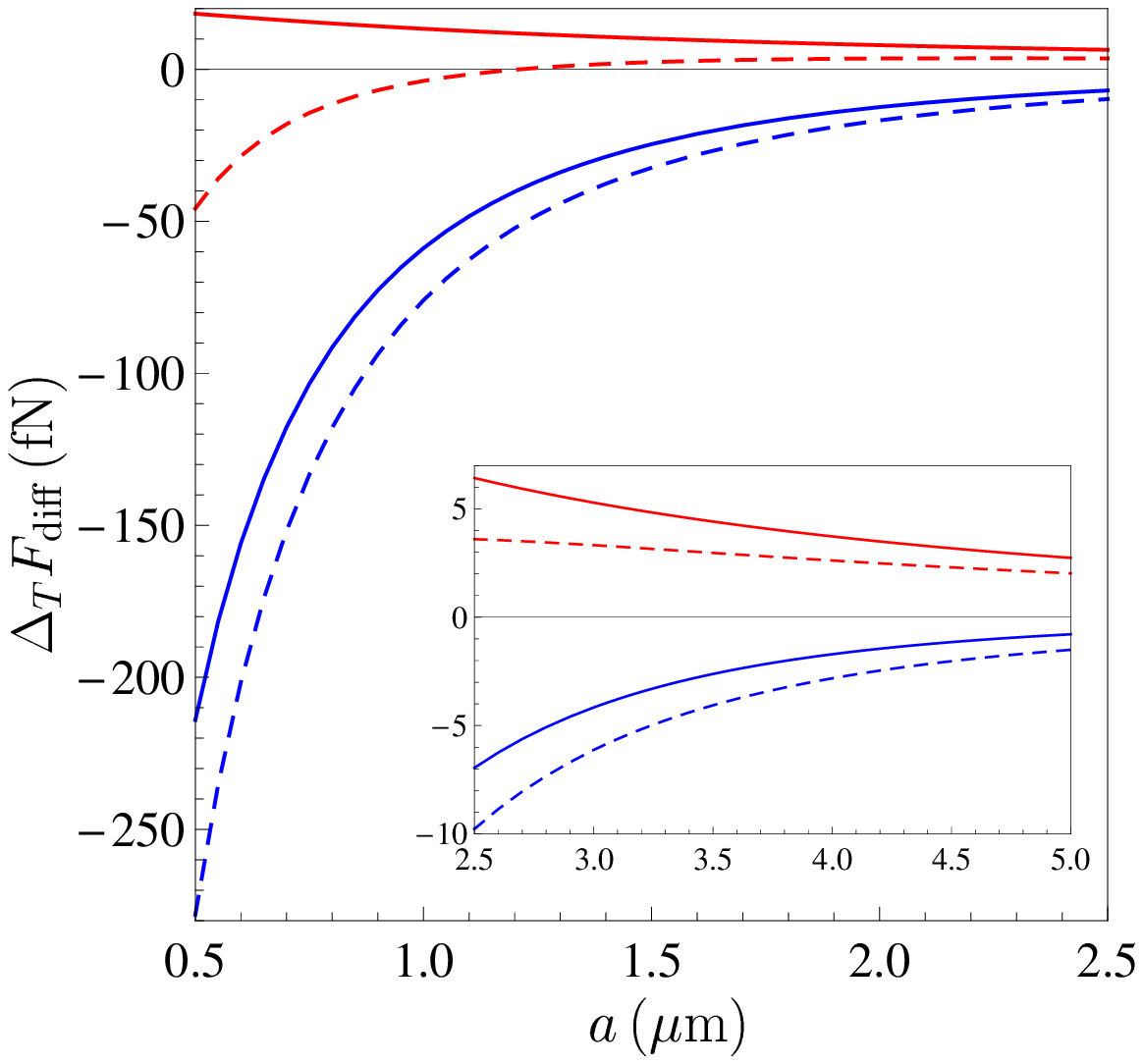}
}
\vspace*{-10cm}
\caption{\label{fg4a}
Thermal corrections to the differential Casimir forces at
$T=300\,$K computed using the plasma model for Au with disregarded and taken
into account
free charge carriers in dielectric materials (top pair of solid
and dashed lines, respectively) and
using the Drude model for Au with disregarded and included
free charge carriers in dielectrics (bottom pair of solid
and dashed lines, respectively) are shown as functions of
separation.
 The region of larger distances is displayed in an inset.
}
\end{figure}

\begin{figure}[b]
\vspace*{-6cm}
\centerline{\hspace*{3.5cm}
\includegraphics{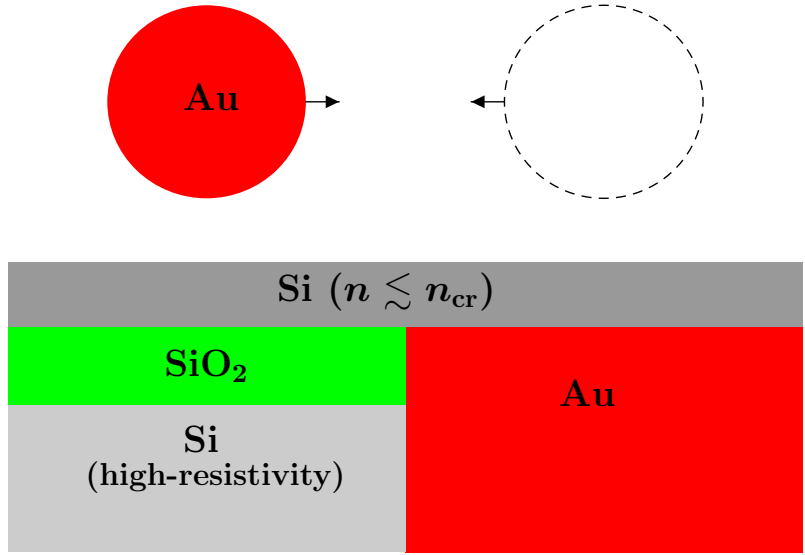}
}
\vspace*{-11.5cm}
\caption{\label{fg5}
The experimental configuration of a Au sphere moving back and
forth above a structured
plate with an additional SiO${}_2$ layer.
The P-doped Si overlayer is in the dielectric state.
The measured quantity is the differential Casimir force $F_{\rm diff}$
between the Au sphere and the two halves of the plate when the sphere bottom
is far away from their boundaries. The figure displays the two extreme
positions of the sphere during its motion.
The size of the sphere is shown not to scale.
}
\end{figure}

The experimental scheme discussed above (see Fig.~\ref{fg1}) allows
for a good discrimination
between the four theoretical approaches and
for a measurement of the thermal effect
in the differential Casimir force. However,
a practical drawback of this scheme is that
it is not easy to produce the structured
plate considered
 in this experiment. Specifically, even a small step between the plate
sections made of high-resistivity Si and Au leads to considerable uncertainties
in the theoretical predictions \cite{44a}. To avoid this problem and
simultaneously increase the relative difference between the different theoretical
predictions, we consider a slightly different experimental scheme. In the modified
setup, the left section of the structured plate below the overlayer contains
an additional layer made of SiO${}_2$ (later denoted as material 4)
 of thickness $D$
followed by bulk high-resistivity Si (previously denoted as
material 3). The modified experimental
scheme is shown in Fig.~\ref{fg5}. An advantage of the modified structured plate
is that it can be manufactured from a commercial wafer of Si grown on an
insulator (SiO${}_2$), i.e., a Si plate with a buried SiO${}_2$ layer,
following the procedure described in Refs.~\cite{2,29}.

The Lifshitz-type formula (\ref{eq3}) for the differential Casimir force
preserves its validity with the replacement
\begin{equation}
R_{\alpha}^{(0,2,3)}(i\zeta_l,y)\to  R_{\alpha}^{(0,2,4,3)}(i\zeta_l,y),
\label{eq26}
\end{equation}
\noindent
where $R_{\alpha}^{(0,2,4,3)}(i\zeta_l,y)$ denotes
the reflection coefficient of the modified left half of
the plate in Fig.~\ref{fg5}, which now consists of the Si
overlayer,
 covering the plate sections made of SiO${}_2$ and
high-resistivity Si:
\begin{equation}
R_{\alpha}^{(0,2,4,3)}(i\zeta_l,y)=
\frac{r_{\alpha}^{(0,2)}(i\zeta_l,y)+R_{\alpha}^{(2,4,3)}(i\zeta_l,y)
e^{-d\sqrt{y^2+(\varepsilon_{2,l}-1)\zeta_l^2}/a}}{1+
r_{\alpha}^{(0,2)}(i\zeta_l,y)R_{\alpha}^{(2,4,3)}(i\zeta_l,y)
e^{-d\sqrt{y^2+(\varepsilon_{2,l}-1)\zeta_l^2}/a}}.
\label{eq27}
\end{equation}
\noindent
Here, the reflection coefficient
$R_{\alpha}^{(2,4,3)}(i\zeta_l,y)$
of the SiO${}_2$ and Si sections of the plate
is given by
\begin{equation}
R_{\alpha}^{(2,4,3)}(i\zeta_l,y)=
\frac{r_{\alpha}^{(2,4)}(i\zeta_l,y)+r_{\alpha}^{(4,3)}(i\zeta_l,y)
e^{-D\sqrt{y^2+(\varepsilon_{4,l}-1)\zeta_l^2}/a}}{1+
r_{\alpha}^{(2,4)}(i\zeta_l,y)r_{\alpha}^{(4,3)}(i\zeta_l,y)
e^{-D\sqrt{y^2+(\varepsilon_{4,l}-1)\zeta_l^2}/a}}.
\label{eq28}
\end{equation}
\noindent
Note that the coefficients $r_{\alpha}^{(k,j)}$ are defined by Eq.~(\ref{eq6})
with the appropriately chosen upper indices.

Now we consider the behavior of the reflection coefficient (\ref{eq27}) at
zero Matsubara frequency. It is easily seen that
irrespective of how one models
 the free charge carriers in the dielectric materials it holds
\begin{equation}
R_{\rm TE}^{(0,2,4,3)}(0,y)=0.
\label{eq29}
\end{equation}
\noindent
If the free charge carriers in the dielectric materials are taken into account,
one obtains
\begin{equation}
R_{\rm TM}^{(0,2,4,3)}(0,y)=1.
\label{eq30}
\end{equation}
\noindent
The case of the TM reflection coefficient with disregarded free charge carriers
in dielectric materials is the most interesting. In this case from Eq.~(\ref{eq27})
one arrives at
\begin{equation}
R_{\rm TM}^{(0,2,4,3)}(0,y)=
\frac{r_{\rm TM,0}^{(0,2)}+R_{\rm TM}^{(2,4,2)}(0,y)
e^{-dy/a}}{1+r_{\rm TM,0}^{(0,2)}R_{\rm TM}^{(2,4,2)}(0,y)
e^{-dy/a}},
\label{eq31}
\end{equation}
\noindent
where $r_{\rm TM,0}^{(0,2)}$ is defined in Eq.~(\ref{eq14}).
The reflection coefficient $R_{\rm TM}^{(2,4,2)}(0,y)$ is given by
\begin{equation}
R_{\rm TM}^{(2,4,2)}(0,y)=r_{\rm TM,0}^{(2,4)}
\frac{1-e^{-Dy/a}}{1-{r_{\rm TM,0}^{(2,4)}}^2e^{-Dy/a}},
\label{eq32}
\end{equation}
\noindent
where
\begin{equation}
r_{\rm TM,0}^{(2,4)}=\frac{\varepsilon_{4,0}^{\rm cor}-
\varepsilon_{2,0}^{\rm cor}}{\varepsilon_{4,0}^{\rm cor}+
\varepsilon_{2,0}^{\rm cor}}
\label{eq33}
\end{equation}
\noindent
and we have used the obvious identity
\begin{equation}
r_{\rm TM,0}^{(4,2)}=-r_{\rm TM,0}^{(2,4)}.
\label{eq34}
\end{equation}

Numerical computations of the differential Casimir force were made by
using Eqs.~(\ref{eq3}), and (\ref{eq26})--(\ref{eq33}).
The dielectric permittivity of SiO${}_2$ with disregarded free charge
carriers, $\varepsilon_{4,l}^{\rm cor}$, is approximated to  high
accuracy by the Ninham-Parsegian representation which takes into account
the effects of electronic and ionic polarization \cite{53,54}.
The oscillator parameters have been determined from a fit to optical
data. The static dielectric permittivity of SiO${}_2$ is
$\varepsilon_{4,0}^{\rm cor}\approx 3.81$. Similar to the case of
high-resistivity Si, an account of free charge carriers in SiO${}_2$
affects the dielectric permittivity only at the zero Matsubara frequency
and results in some dc conductivity. Taking into consideration that due to
Eq.~(\ref{eq29}) the coefficient $R_{\rm TE}^{(0,2,4,3)}(0,y)$ does not
depend on the account or neglect of free charge carriers, their impact is
determined by the coefficient
$R_{\rm TM}^{(0,2,4,3)}(0,y)$
in accordance to Eqs.~(\ref{eq30}) and (\ref{eq31}).

\begin{figure}[b]
\vspace*{-4cm}
\centerline{\hspace*{3.5cm}
\includegraphics{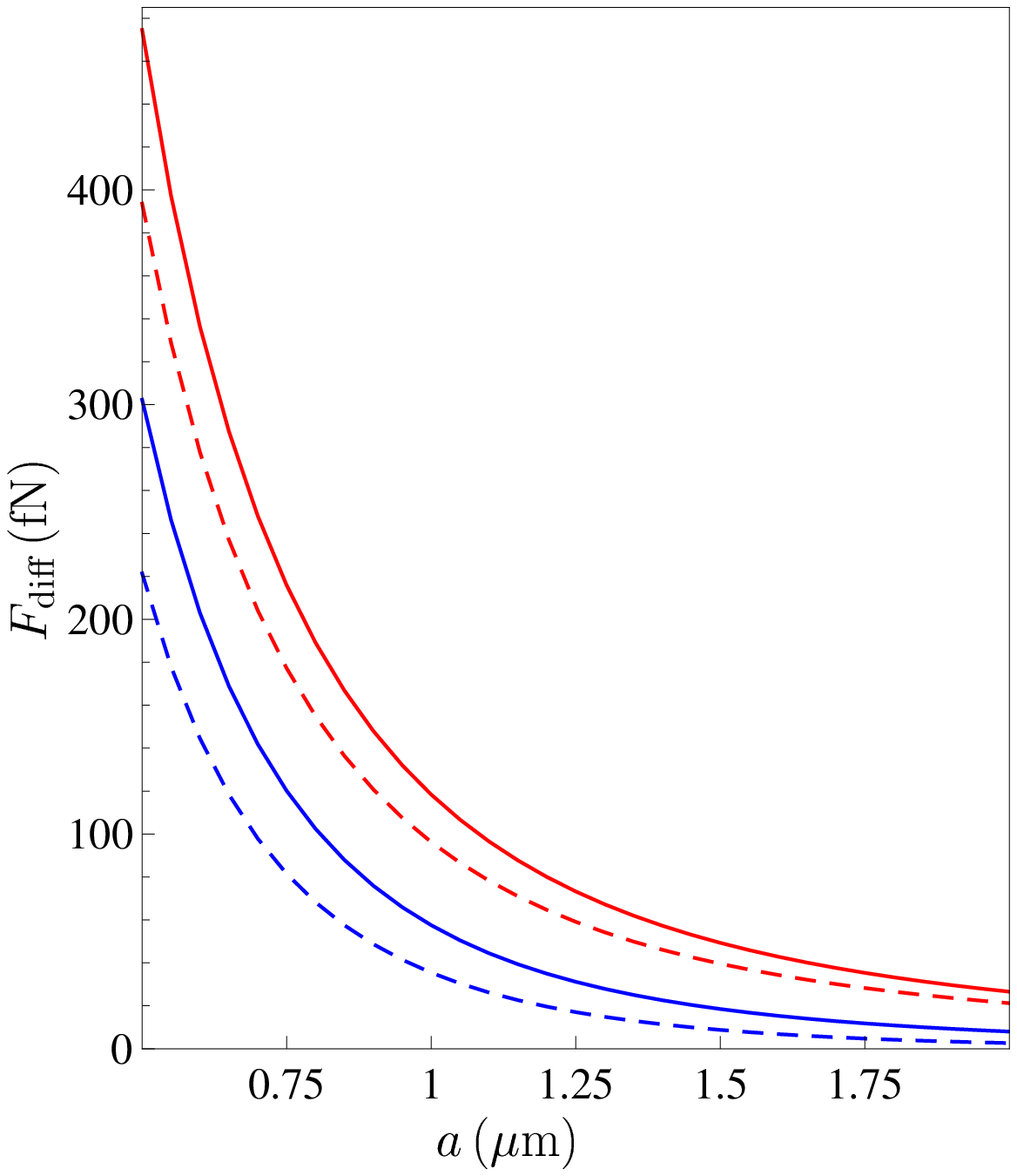}
}
\vspace*{-10cm}
\caption{\label{fg6}
Differential Casimir forces computed at $T=300\,$K
in the configuration with additional SiO${}_2$ layer (Fig.~\ref{fg5})
using the plasma model for Au with disregarded and taken into account
free charge carriers in dielectric materials (top pair of solid
and dashed lines, respectively) and
using the Drude model for Au with disregarded and included
free charge carriers in dielectrics (bottom pair of solid
and dashed lines, respectively) are shown as functions of
separation.
}
\end{figure}

Computations have been performed at $T=300\,$K taking for the thickness of
the P-doped overlayer the value of
$d=200\,$nm and the value of $D=400\,$nm for the
thickness of the SiO${}_2$ layer.
 The computational results for the differential Casimir force
are presented in Fig.~\ref{fg6} as  functions of separation by the four
lines corresponding to the four theoretical approaches
(the top pair of solid and dashed lines is computed using the plasma model
for Au with disregarded and taken into account free charge carriers in all
dielectric materials, respectively, whereas
the bottom pair of lines is obtained using the Drude model
for Au and the same options for free charges in dielectrics).

As is seen in Fig.~\ref{fg6}, all four theoretical approaches lead to widely
distinct differential Casimir forces, which can be discriminated
experimentally with certainty within the separation region from 0.5 to
$2\,\mu$m. The computational results for the differential Casimir forces
computed using the four theoretical approaches are presented in Table~I over
the separation region from 0.5 to $3\,\mu$m. The first column contains
the value of separation. The second and third columns list the values of
$F_{\rm diff}^{p}$ and $F_{\rm diff}^{p,dc}$ computed using the plasma model
for Au with disregarded and taken into account free charge carriers of
dielectric materials, respectively. The fourth and fifth columns
present the values of
$F_{\rm diff}^{D}$ and $F_{\rm diff}^{D,dc}$ obtained using the Drude model
for Au with disregarded and taken into account free charge carriers of
dielectric materials, respectively.

According to Table~I, at $a=0.5\,\mu$m the differences between the predicted
values $F_{\rm diff}^{p}$ and $F_{\rm diff}^{p,dc}$, between
$F_{\rm diff}^{p,dc}$ and $F_{\rm diff}^{D}$, and between
$F_{\rm diff}^{D}$ and $F_{\rm diff}^{D,dc}$ are equal to
80.22, 92.06, and 80.17\,fN, respectively.
At $a=1.0$ and $1.6\,\mu$m the same differences are equal  to
21.98, 38.86, and 21.98\,fN and 8.49, 19.06, and 8.48\,fN, respectively.
All these values are far in excess of the experimental sensitivity.

If the plasma model for Au is used in computations, the relative deviation of
the differential Casimir force obtained with disregarded free charge carriers
in dielectrics from that found with taken free charge carriers into account is
\begin{equation}
\delta F_{\rm diff}^{p}\equiv
\frac{F_{\rm diff}^{p}-F_{\rm diff}^{p,dc}}{F_{\rm diff}^{p,dc}}.
\label{eq35}
\end{equation}
\noindent
{}From Table~I one obtains that $\delta F_{\rm diff}^{p}=20.3\%$, 22.8\%, and
24.7\% at separations $a=0.5$, 1.0, and $1.6\,\mu$m, respectively.
The analogous deviation between the next two theoretical approaches defined as
\begin{equation}
\delta F_{\rm diff}^{p,dc;D}\equiv
\frac{F_{\rm diff}^{p,dc}-F_{\rm diff}^{D}}{F_{\rm diff}^{D}}
\label{eq36}
\end{equation}
\noindent
is equal to 30.4\%, 67.6\%, and 124.4\% at the same respective separations.
Finally, the relative deviation of $F_{\rm diff}^{D}$ from $F_{\rm diff}^{D,dc}$
is given by
\begin{equation}
\delta F_{\rm diff}^{D}\equiv
\frac{F_{\rm diff}^{D}-F_{\rm diff}^{D,dc}}{F_{\rm diff}^{D,dc}}.
\label{eq37}
\end{equation}
\noindent
At separation distances $a=0.5$, 1.0, and $1.6\,\mu$m one obtains from
Table~I $\delta F_{\rm diff}^{D}=36.1\%$, 61.8\%, and 124.0\%, respectively.
As to the relative deviation between the extreme two approaches,
namely the one  which
disregards both the relaxation properties of electrons in metals and free charge
carriers in dielectrics and the other which instead takes both into account,
it is defined as
\begin{equation}
\delta F_{\rm diff}^{p;D,dc}\equiv
\frac{F_{\rm diff}^{p}-F_{\rm diff}^{D,dc}}{F_{\rm diff}^{D,dc}}.
\label{eq38}
\end{equation}
\noindent
{}From Table~I one obtains that $\delta F_{\rm diff}^{p;D,dc}=113.6\%$, 233\%, and
526.8\% at the same respective separations.
One can conclude that all the above relative deviations are sufficiently
large for experimental discrimination between different theoretical approaches and
all of them quickly increase with increasing separation.

\begin{figure}[b]
\vspace*{-6cm}
\centerline{\hspace*{3.5cm}
\includegraphics{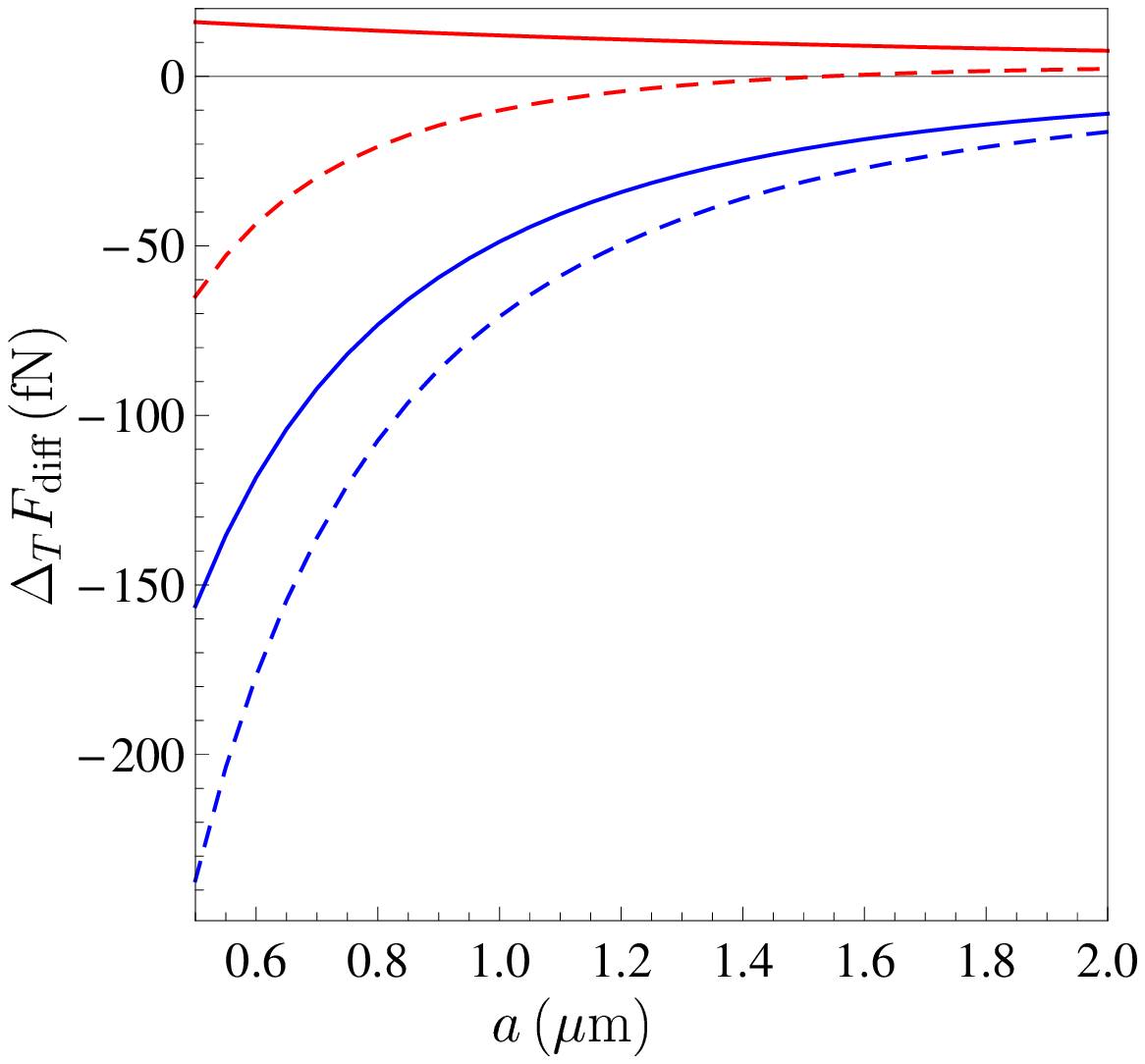}
}
\vspace*{-10cm}
\caption{\label{fg8}
Thermal corrections to the differential Casimir forces computed at
$T=300\,$K
in the configuration with additional SiO${}_2$ layer
using the plasma model for Au with disregarded and taken
into account
free charge carriers in dielectric materials (top pair of solid
and dashed lines, respectively) and
using the Drude model for Au with disregarded and included
free charge carriers in dielectrics (bottom pair of solid
and dashed lines, respectively) are shown as functions of
separation.
}
\end{figure}

We have also computed the thermal correction (\ref{eq26a}) in the
differential Casimir force
in the configuration of Fig.~\ref{fg5} in the framework of the
four theoretical approaches described above.
The computational results at $T=300\,$K are shown in
Fig.~\ref{fg8}
as functions of separation.
The top pair of solid and dashed lines is obtained by
using the plasma model for Au
with disregarded and taken into account
free charge carriers in dielectrics, respectively.
The bottom pair of solid and dashed lines is computed by
means of the Drude model for Au
with respective neglect and inclusion of
free charge carriers in dielectric materials.
As can be seen in Fig.~\ref{fg8}, over the separation range
from 0.5 to $2\,\mu$m the thermal corrections predicted by
the four theoretical approaches are significantly different
and can be easily discriminated
from each other
taking into account the
experimental sensitivity. Thus, at separations $a=0.5$,
1.0, 1.5, and $2.0\,\mu$m the thermal correction
using the plasma model with omitted conductivity of
dielectrics, $\Delta_TF_{\rm diff}^{p}$, is equal to
16.0, 12.1, 9.4, and 7.6\,fN,
respectively.
In a similar way, for the remaining three theoretical
approaches one has
\begin{eqnarray}
&&
\Delta_TF_{\rm diff}^{p,dc}=
-64.9,{\ }-10.0,{\ }-0.3,{\ }2.2
\,\mbox{fN},
\nonumber \\
&&
\Delta_TF_{\rm diff}^{D}=
-156.3,{\ }-48.8,{\ }-21.4,{\ }-11.0
\,\mbox{fN},
\label{38a} \\
&&
\Delta_TF_{\rm diff}^{D,dc}=
-237.2,{\ }-70.8,{\ }-31.1,{\ }-16.4
\,\mbox{fN}
\nonumber
\end{eqnarray}
\noindent
at the same respective separations.

\section{Estimation of theoretical errors}

Here, we present an estimation of errors inherent to the above computations.
As was mentioned in Sec.~III, a conservative estimation of the minimum detectable
force in experiments of this type is
$\Delta F_{\rm diff}^{\rm expt}=1\,$fN.
The total theoretical error in the determination of $F_{\rm diff}$ is contributed
by several independent components and depends on separation.

We begin with a possible error in the concentration of free charge carriers
in the P-doped overlayer $\delta n=5\%$ (see Fig.~\ref{fg5}).
This error does not influence the computed values of $F_{\rm diff}^{D}$ and
$F_{\rm diff}^{p}$ when the free charge carriers in dielectric materials are
disregarded, but leads to the following errors if the free charge carriers
are taken into account
\begin{eqnarray}
&&
\delta_nF_{\rm diff}^{D,dc}(a_1)\approx 0.14\%,
\quad
\delta_nF_{\rm diff}^{D,dc}(a_2)\approx 0.15\%,
\nonumber \\
&&
\delta_nF_{\rm diff}^{p,dc}(a_1)\approx 0.08\%,
\quad
\delta_nF_{\rm diff}^{p,dc}(a_2)\approx 0.02\%,
\label{eq39}
\end{eqnarray}
\noindent
where we present the values of errors at $a_1=0.5\,\mu$m and
$a_2=2\,\mu$m. These results were obtained by repeating the computations of
Sec.~IV with $\tilde{n}=n\pm 0.05n$.

The second error source in theoretical values of the differential Casimir
force is  uncertainty in the thickness of the SiO${}_2$ layer.
For a commercial Si wafer $\Delta D=2\,$nm leading to $\delta D=0.5\%$ in our
case. This leads to the following relative errors in our computational results:
\begin{eqnarray}
&&
\delta_DF_{\rm diff}^{D}(a_1)\approx 0.05\%,
\quad
\delta_DF_{\rm diff}^{D}(a_2)\approx 0.075\%,
\nonumber \\
&&
\delta_DF_{\rm diff}^{p}(a_1)\approx 0.03\%,
\quad
\delta_DF_{\rm diff}^{p}(a_2)\approx 0.05\%,
 \label{eq40} \\
&&
\delta_DF_{\rm diff}^{D,dc}(a_1)\approx 0.04\%,
\quad
\delta_DF_{\rm diff}^{D,dc}(a_2)\approx 0.06\%,
\nonumber \\
&&
\delta_DF_{\rm diff}^{p,dc}(a_1)\approx 0.02\%,
\quad
\delta_DF_{\rm diff}^{p,dc}(a_2)\approx 0.008\%.
\nonumber
\end{eqnarray}

The third source of theoretical errors is the error in the thickness of Si overlayer
$\delta d=2\,$nm, i.e., $\delta d=1\%$ resulting in
\begin{eqnarray}
&&
\delta_dF_{\rm diff}^{D}(a_1)\approx 1.3\%,
\quad
\delta_dF_{\rm diff}^{D}(a_2)\approx 0.5\%,
\nonumber \\
&&
\delta_dF_{\rm diff}^{p}(a_1)\approx 1.05\%,
\quad
\delta_dF_{\rm diff}^{p}(a_2)\approx 0.5\%,
 \label{eq41} \\
&&
\delta_dF_{\rm diff}^{D,dc}(a_1)\approx 1.6\%,
\quad
\delta_dF_{\rm diff}^{D,dc}(a_2)\approx 1.1\%,
\nonumber \\
&&
\delta_dF_{\rm diff}^{p,dc}(a_1)\approx 1.1\%,
\quad
\delta_dF_{\rm diff}^{p,dc}(a_2)\approx 0.3\%.
\nonumber
\end{eqnarray}

All these errors are random quantities characterized by a uniform distribution.
In this case the total relative error is obtained by a summation of the
respective errors (\ref{eq39})--(\ref{eq41}) \cite{2,55}
\begin{eqnarray}
&&
\delta_{\rm tot}F_{\rm diff}^{D}(a_1)\approx 1.4\%,
\quad
\delta_{\rm tot}F_{\rm diff}^{D}(a_2)\approx 0.6\%,
\nonumber \\
&&
\delta_{\rm tot}F_{\rm diff}^{p}(a_1)\approx 1.1\%,
\quad
\delta_{\rm tot}F_{\rm diff}^{p}(a_2)\approx 0.6\%,
 \label{eq42} \\
&&
\delta_{\rm tot}F_{\rm diff}^{D,dc}(a_1)\approx 1.8\%,
\quad
\delta_{\rm tot}F_{\rm diff}^{D,dc}(a_2)\approx 1.3\%,
\nonumber \\
&&
\delta_{\rm tot}F_{\rm diff}^{p,dc}(a_1)\approx 1.2\%,
\quad
\delta_{\rm tot}F_{\rm diff}^{p,dc}(a_2)\approx 0.3\%,
\nonumber
\end{eqnarray}

There is one more theoretical uncertainty originating from errors in the
dielectric permittivities. At separations in the micrometer range, errors
in the optical data do not lead to noticeable errors in the differential
Casimir force. However, depending on the properties of a specific Au sample
and its preparation process, the minimum possible value of the plasma
frequency of Au was found to be $\omega_{p,1}^{\min}\approx 6.8\,$eV \cite{56,57}.
The corresponding value of the relaxation parameter is
$\gamma_1^{\min}\approx 0.02\,$eV. In fact, different values of $\omega_{p,1}$
in the interval from 6.8 to 9.0\,eV cannot be considered as random values of
the measured quantity because they correspond to different samples. It would be
preferable to determine the value of $\omega_p$ for the specific sample used in
measurements of the Casimir force as it was done in Refs.~\cite{19,20}.
However, for illustrative purposes, here we present the relative deviations of the
value of $F_{\rm diff}$ which would be obtained upon using
for the plasma frequency the value $\omega_{p,1}^{\min}$
instead of the value $\omega_{p,1}=9.0\,$eV that was used above:
\begin{eqnarray}
&&
\delta_{\omega_p}F_{\rm diff}^{D}(a_1)\approx -0.7\%,
\quad
\delta_{\omega_p}F_{\rm diff}^{D}(a_2)\approx -0.03\%,
\nonumber \\
&&
\delta_{\omega_p}F_{\rm diff}^{p}(a_1)\approx -1.5\%,
\quad
\delta_{\omega_p}F_{\rm diff}^{p}(a_2)\approx -0.6\%,
 \label{eq43} \\
&&
\delta_{\omega_p}F_{\rm diff}^{D,dc}(a_1)\approx -1.3\%,
\quad
\delta_{\omega_p}F_{\rm diff}^{D,dc}(a_2)\approx -0.8\%,
\nonumber \\
&&
\delta_{\omega_p}F_{\rm diff}^{p,dc}(a_1)\approx -0.6\%,
\quad
\delta_{\omega_p}F_{\rm diff}^{p,dc}(a_2)\approx -0.85\%,
\nonumber
\end{eqnarray}

Some other effects which may influence the value of the differential force
are the surface roughness and electrostatic patches. However, as it was
noticed in Refs.~\cite{37,37a,44a,44b,65}, the effect of patches
is strongly suppressed in
the differential force, whereas the effect of roughness is negligible in the
micrometer separation range \cite{2,3,37aa}. Note that in comparison between the
measurement data and theory in the proposed experiment one should correct the
theoretical values in Table~I for deviations from the proximity force
approximation used in Eq.~(\ref{eq3}). This correction can be estimated
as of about
--0.2\% of $F_{\rm diff}$ \cite{21,46,47,48}.

{}From Eqs.~(\ref{eq42}) and (\ref{eq43}) it is seen that even taking into
account all possible errors and uncertainties the theoretical predictions of
the four
approaches discussed above differ considerably and can be easily discriminated by
performing the proposed experiment.

\section{Conclusions and discussion}

In the foregoing, we have suggested a universal setup allowing to directly
measure the thermal effect in the Casimir force
and to determine the role of free charge carriers
in both metallic and dielectric materials, by a single experiment. According to
our results, these aims can be achieved by measuring the differential Casimir
force between a Au-coated sphere moving back and fourth above a structured
plate covered with a conductive overlayer. One half of the plate
is
made of high-resistivity dielectric materials, while its other half
is made of Au.
The main novel feature of this setup is that the conductive overlayer
is made of a doped semiconductor (P-doped Si in our case), whose
concentration of free charge carriers is only slightly below the critical one,
where the dielectric-to-metal phase transition occurs. Thus, the overlayer is
made of a dielectric material possessing at room temperature a rather high
conductivity, allowing for the
electrostatic calibration required in precise
measurements of the Casimir force.

We have considered two versions of the structured plate where the dielectric
section is made either of bulk high-resistivity Si or of a layer of SiO${}_2$
followed by bulk high-resistivity Si. In both cases the differential Casimir force
was calculated over the separation region from $0.5\,\mu$m to a few
micrometers, i.e., in the domain where thermal effects determined by the
zero-frequency term in the Lifshitz formula contribute considerably.

Computations have been performed within the four theoretical approaches discussed
in the literature, i.e., Au at low frequencies is described by the plasma model
and the free charge carriers in all dielectric materials (including the P-doped Si)
are either disregarded or taken into account, or
Au at low frequencies is described by the Drude model
and the free charge carriers in all dielectric materials
are again either disregarded or taken into account.
We have also calculated the dependence of all differential forces on the thickness
of the P-doped Si overlayer, determined the thermal contribution to
the differential Casimir force at different separations and estimated all
related errors and uncertainties.

The obtained results allow to conclude that all the four theoretical approaches
lead to significantly different values of the differential Casimir force in the
micrometer separation range, which are many times larger than both the
experimental and theoretical errors. Thus, the proposed experiment
is capable to provide
an unequivocal confirmation to one of the above theoretical models and rule
out the other three. Our calculation results show that the thermal effect in
the differential Casimir force  is up to two and one orders of magnitude
larger than the minimum detectable signal when the Drude and plasma models
are used, respectively. Because of this, the proposed experiment can not
only lead to
confirmation of one of the models, but it also allows
reliable measurement of the thermal
contribution to the observed signal.

One may guess which of the four above theoretical approaches has the best
chances to be confirmed.
As discussed in Sec.~I, previous experiments with metallic test bodies
\cite{17,18,19,20,21,22,23,38} are in agreement with the plasma model and
exclude the  Drude model, whereas only the experiment \cite{25} leads to
the opposite conclusion. Concurrently, several other experiments performed
with dielectric test bodies \cite{28,29,30,31,32} are in agreement with
theory disregarding the free charge carriers and exclude theory taking the free
charge carriers into account. On this basis, it is reasonable to expect that
the proposed universal experiment may confirm the theoretical prediction
$F_{\rm diff}^{p}$ obtained using the plasma model at low frequencies for Au
with disregarded free charge carriers in dielectric materials.
It should be underlined, however, that almost all of the abovementioned
experiments (with exception of only Refs.~\cite{25,28}) are most precise at
separations below $0.5\,\mu$m and use quite distinct experimental setups.
This allows to conclude that the proposed universal experiment, which is
capable to determine the role of free charge carriers in the Casimir force
between any materials and measure the thermal effect in the micrometer
separation range, will bring challenging results for the theory of
electromagnetic fluctuations.

\section*{Acknowledgments}

The work of V.M.M. was partially supported by the Russian Government
Program of Competitive Growth of Kazan Federal University.

\newpage
\begingroup
\squeezetable
\begin{table}
\caption{The values of the differential Casimir force
computed at $T=300\,$K in the configuration with additional
SiO${}_2$ section using the plasma model for Au with disregarded and
taken into account free charge carriers in dielectric materials
(columns 2 and 3, respectively) and using the Drude model for Au with
disregarded and included free charge carriers in dielectrics
(columns 4 and 5, respectively) are shown at different separations
(column 1).
}
\begin{ruledtabular}
\begin{tabular}{crrrr}
$a\,(\mu\mbox{m})$ & $F_{\rm diff}^{p}\,$(fN) &$F_{\rm diff}^{p,dc}\,$(fN) &
$F_{\rm diff}^{D}\,$(fN) &$F_{\rm diff}^{D,dc}\,$(fN)\\
\hline
0.5 & 474.57~~ & 394.35~~ & 302.29~~ & 222.12~~ \\
0.6 & 336.31~~ & 278.28~~ & 202.96~~ & 144.96~~ \\
0.7 & 248.13~~ & 204.43~~ & 141.86~~ & 98.18~~ \\
0.8 & 189.07~~ & 155.13~~ & 102.39~~ & 68.46~~ \\
0.9 & 147.94~~ & 120.90~~ & 75.88~~ & 48.86~~ \\
1.0 & 118.36~~ & 96.38~~ & 57.52~~ & 35.54~~ \\
1.2 & 80.00~~ & 64.72~~ &34.94~~ & 19.67~~ \\
1.4 & 57.27~~ & 46.10~~ & 22.56~~ & 11.40~~ \\
1.6 & 42.87~~ & 34.38~~ & 15.32~~ & 6.84~~ \\
1.8 & 33.26~~ & 26.61~~ & 10.86~~ & 4.21~~ \\
2.0 & 26.56~~ & 21 22~~ & 7.98~~ & 2.64~~ \\
2.2 & 21.72~~ & 17.34~~ & 6.07~~ & 1.68~~ \\
2.4 & 18.11~~ & 14.46~~ & 4.74~~ & 1.09~~ \\
2.6 & 15.36~~ & 12.26~~ & 3.80~~ & 0.71~~ \\
2.8 & 13.20~~ & 10.55~~ & 3.12~~ & 0.46~~ \\
3.0 & 11.48~~ & 9.19~~ & 2.61~~ & 0.31~~
\end{tabular}
\end{ruledtabular}
\end{table}
\endgroup
\newpage
\end{document}